\documentclass{aa}

\usepackage{graphicx}
\usepackage{txfonts}
\begin{document}

\title{Non-resonant Alfv\'{e}nic instability \\
activated by high temperature of ion beams \\
in compensated-current astrophysical plasmas}

\titlerunning{Non-resonant Alfv\'{e}nic instability activated by high beam
temperature}

   \author{P. Malovichko
          \inst{1}
          \and
          Y. Voitenko
          \inst{2}
          \and
          J. De Keyser
          \inst{2}
          }

   \institute{Main Astronomical Observatory, NASU, Kyiv, Ukraine\\
              \email{malovichp@i.ua}
         \and
             Solar-Terrestrial Centre of Excellence,
             Royal Belgian Institute for Space Aeronomy,
             Ringlaan 3, 1180 Brussels, Belgium\\
             \email{voitenko@oma.be}
             }

   \date{Received 04 August 2017; accepted 23 April 2018}


  \abstract
   {Compensated-current systems are established in response to hot ion beams in
terrestrial foreshock regions, around supernova remnants, and in other space
and astrophysical plasmas.}
   {We study a non-resonant reactive instability of Alfv\'{e}n waves (AWs)
propagating quasi-parallel to the background magnetic field $\mathbf{B}_{0}$
in such systems.  }
   {The instability is investigated analytically in the framework of kinetic
theory applied to the hydrogen plasmas penetrated by hot proton beams. }
   {The instability arises at parallel wavenumbers $k_{z}$ that are
sufficiently large to demagnetize the beam ions, $k_{z}V_{Tb}/\omega
_{Bi}\gtrsim $ $1$ (here $V_{Tb}$ is the beam thermal speed along $\mathbf{B}%
_{0}$ and $\omega _{Bi}$ is the ion-cyclotron frequency). The Alfv\'{e}n
mode is then made unstable by the imbalance of perturbed currents carried by
the magnetized background electrons and partially demagnetized beam ions.
The destabilizing effects of the beam temperature and the temperature
dependence of the instability threshold and growth rate are demonstrated for
the first time. The beam temperature, density, and bulk speed are all
destabilizing and can be combined in a single destabilizing factor $\alpha
_{b}$ triggering the instability at {$\alpha _{b}>$ $\alpha _{b}^{\mathrm{thr%
}}$}, where the threshold value varies in a narrow range $2.43\leq $ $\alpha
_{b}^{\mathrm{thr}}\leq $ $4.87$. New analytical expressions for the
instability growth rate and its boundary in the parameter space are obtained
and can be directly compared with observations. Two applications to
terrestrial foreshocks and foreshocks around supernova remnants are shortly
discussed. In particular, our results suggest that the ions reflected by the
shocks around supernova remnants can drive stronger instability than the
cosmic rays.}
   {}


   \maketitle
%

\section{Introduction}

Diluted ion beams propagating along the background magnetic field $\mathbf{B}%
_{0}$ are widespread in space and astrophysical plasmas, including solar
wind \citep[][and references
therein]{Marsch2006}, terrestrial foreshocks
\citep[][and references
therein]{Paschmann1981}, supernova remnants
\citep[][and references
therein]{Bell2005}, and many other astrophysical environments
\citep[][and references
therein]{Zweibel2010}. As the plasmas are typically quasi-neutral, the
background electrons tend to follow the beam ions compensating their
current. Depending on particular settings, the compensating currents can
also be provided by other plasma components, like co-streaming electron
beams injected simultaneously with the ion beams. Plasma instabilities
developing in such compensated-current systems not only regulate the plasma
and beam parameters keeping them close to the marginally unstable states,
but can also be important sources for the background plasma heating,
energetic particles acceleration, and amplification of the background
magnetic field.

Plasma waves in the compensated-current systems can be driven unstable by
resonant \citep{Duijveman1981,Gary1985,Voitenko1990} and non-resonant %
\citep{Winske1984,Bell2004,Achterberg2013} wave-particle interactions.
Resonant kinetic instabilities of various wave modes, driven by the beam
ions, have been studied extensively in the past. Parallel-propagating Alfv%
\'{e}n and fast waves have been found to be most unstable for the beam
velocities larger than few Alfv\'{e}n velocities
\citep[e.g.][and references
therein]{Gary2005,Marsch2006}. Concurrent instabilities of oblique (kinetic)
Alfv\'{e}n waves come into play at lower (but still super-Alfv\'{e}nic) beam
velocities \citep{Voitenko1998}.

The mentioned above instabilities can be driven by the beam ions %
\citep{Sentman1981,Winske1984,Gary1985} or by the electron return currents %
\citep[][and references therein]{Winske1984,Bell2004,Chen2012}. The
non-compensated electron currents flowing along $\mathbf{B}_{0}$, may also
drive both the resonant \citep[][]{Voitenko1995} and non-resonant %
\citep[][]{Malovichko1992,Malovichko2007} instabilities of Alfven waves. The
simplest case of purely parallel propagating Alfv\'{e}n waves has been
considered in application to the current-carrying coronal loops %
\citep[][]{Malovichko1992}, where these waves appeared to be always
unstable. Later on, the analysis has been extended by accounting for the
oblique propagation \citep[][]{Malovichko2007} and the currents curried by
low-density beams \citep[][]{Malovichko2010}, and applied to the terrestrial
magnetosphere and coronal loops.

Self-consistent modifications of the background magnetic field by the
electric currents, neglected in %
\citep[][]{Malovichko1992,Malovichko2007,Voitenko1995}, may reduce or even
stabilize current instabilities. This issue does not concern instabilities
developing in compensated-current systems. Such systems, formed around
supernova remnants by high-energy streaming cosmic rays (CRs), have been
studied by \cite{Bell2004}, who found a new non-resonant Alfv\'{e}nic
instability (Bell instability thereafter). Since then, the Bell instability
and its modifications attracted a lot of interest
\citep[see e.g.][and references
therein]{Amato2009,Bret2009,Zweibel2010,Schure2012,Achterberg2013,Kobzar2017}%
. Following \cite{Bell2004}, the primary focus has been on the unstable
modes with finite $k_{z}\bar{V}_{bz}/\omega _{Bi}$\ propagating along $B_{0}$%
\ ($\bar{V}_{bz}$\ is a characteristic velocity of the beam ions along the
mean magnetic field $B_{0}\parallel z$, $k_{z}$ is the parallel wavenumber,
and $\omega _{Bi}$\ is the ion-cyclotron frequency). Compensated currents
can also drive an oblique Alfv\'{e}n instability \citep[][]{Malovichko2014},
for which the perpendicular wave dispersion due to finite $k_{\perp
}V_{Tb\perp }/\omega _{Bi}$ is essential ($k_{\perp }$ and $V_{Tb\perp }$
are the perpendicular wavenumber and beam thermal velocity in the plane $%
\perp \mathbf{B}_{0}$).

Other electrostatic and electromagnetic instabilities may develop in
compensated-current systems \citep[see e.g.][and references
therein]{Gary2005,Bret2009,Brown2013,Marcowith2016}. What wave modes grow
fastest critically depends on the beam and plasma parameters. In the case of
cold diluted proton beams propagating along $B_{0}$, the electrostatic
two-stream and Buneman instabilities are much faster than the
electromagnetic Alfv\'{e}nic instabilities \citep[see e.g. Fig. 44
by][]{Bret2010}. Nevertheless, as is noted by \cite{Bret2010}, these
electrostatic instabilities are quickly saturated, and then electromagnetic
Alfv\'{e}nic/Bell instabilities come into play. In the hot beam/plasma
systems, where the two-stream/Buneman instabilities cannot develop, the
electromagnetic Alfv\'{e}nic/Bell instabilities dominate.

The Bell instability has the maximum growth rate $\gamma _{\mathrm{Bell}%
}\simeq 0.5\bar{j}_{b}\omega _{Bi}$, where $\bar{j}_{b}=n_{b}V_{b}/\left(
n_{0}V_{A}\right) $ is the beam current normalized by the Alfv\'{e}n
current. This maximum is attained at the parallel wavenumber $\left\vert k_{z%
\mathrm{m}}\right\vert V_{A}/\omega _{Bi}=$ $0.5\bar{j}_{b}$ and the
perpendicular wavenumber $k_{\perp }=0$. These expressions are exactly the
same as for the instability studied earlier by \cite{Winske1984} in
application to the terrestrial foreshock. The difference is that the role of
$\bar{V}_{bz}$ in the setting considered by \cite{Winske1984} is played by
the bulk velocity of the beam $V_{b}$ rather than the large velocity spread
of CRs. Both the Winske-Leroy and Bell instabilities grow fastest when the
wave vector $k$\ is parallel to $B_{0}$; they are physically the same
instability that can be named the compensated-current parallel instability
(CCPI).

The physical mechanism of CCPI is related to the fact that for sufficiently
small parallel wavelengths and sufficiently high $\bar{V}_{bz}$, the beam
protons become partially demagnetized (unfrozen off the perturbed magnetic
field). The demagnetization reduces the beam contribution to the fluctuating
currents $\delta \mathbf{j}\perp \mathbf{B}_{0}$ flowing along the (twisted)
perturbed magnetic field lines, whereas the electron currents remain
magnetized thus providing the non-compensated fluctuating transversal
currents. These currents amplify the initial perturbations via the positive
feedback loop giving rise to CCPI. This kind of instabilities is sometimes
called reactive.

Surprisingly, despite of its importance in astrophysical applications, the
CCPI theory is still poorly developed. Many important properties of the
instability (the wavenumber dependence of the instability growth rate,
behavior of the maximum growth rate in the parameter space, instability
boundaries in the parameter spaces, etc.) have not been fully investigated.
In the present paper, we study CCPI of Alfv\'{e}n waves in more detail in
the framework of kinetic theory. We consider a simple model of the
compensated-current system where the hydrogen plasma is penetrated by the
low-density proton beam and the beam current and charge are compensated by
the background electrons. Despite of its simplicity, this model is
applicable to the reactive CCPI driven by compensated currents in many space
and astrophysical environments.

\section{Plasma model and dispersion equation for Alfv\'{e}n waves}

We consider a three-component plasma consisting of the background steady ion
component ($i$), the low-density ion beam ($b$) propagating with velocity $%
V_{b}$\ along $z\parallel B_{0}$, and the electron component ($e$) providing
the neutralizing current\ and charge:
\begin{equation}
n_{e}V_{e}=n_{b}V_{b};  \label{j=0}
\end{equation}%
\begin{equation}
n_{e}=n_{i}+n_{b}\equiv n_{0}.  \label{q=0}
\end{equation}%
We assume here that the beam ions ($b$) and the background ions ($i$) are
protons. All plasma components are modeled by the shifted Maxwellian
velocity distributions
\begin{equation}
f_{0s}=\frac{n_{s}}{(2\pi )^{3/2}V_{Ts}^{3}}\exp \left( -\frac{v_{\bot }^{2}%
}{V_{Ts}^{2}}-\frac{(v_{z}-V_{s})^{2}}{V_{Ts}^{2}}\right) ,  \label{f0}
\end{equation}%
where $n_{s}$,\ $V_{s}$, $V_{Ts}=\sqrt{T_{s}/m_{s}}$, $T_{s}$ and $m_{s}$
are the mean number density, parallel bulk velocity, thermal velocity,
temperature and particle mass of the plasma specie $s$, and $\mathbf{v=}%
\left( v_{x},v_{y,}v_{z}\right) $ - velocity-space coordinates. The
subscripts $z$ and $\perp $ indicate directions parallel and perpendicular
to $\mathbf{B}_{0}$. The plasma model defined by (\ref{j=0}-\ref{f0}) has
been extensively used in the past \citep[see e.g.][and references
therein]{Gary2005}. The neutralizing current can also be provided by the
co-propagating electron beam \citep[see e.g.][and references
therein]{Zweibel2010}, which however does not alter the reactive CCPI for
low-density ion beams $n_{b}\ll n_{0}$ \citep{Amato2009}.

The nontrivial solutions to the Maxwell-Vlasov set of equations exist if the
perturbation wave frequency $\omega $ and the wave vector $\mathbf{k}%
=(k_{x},k_{y},k_{z})$ satisfy the following dispersion equation \citep[see
e.g.][]{Alexandrov1984}:
\begin{equation}
\left\vert k^{2}\delta _{ij}-k_{i}k_{j}-\frac{\omega ^{2}}{c^{2}}\varepsilon
_{ij}\right\vert =0,  \label{det}
\end{equation}%
where $\varepsilon _{ij}$ is the dielectric tensor, and $\delta _{ij}$ is
the Kronecker's delta-symbol. For the parallel-propagating modes with $%
k_{x}=k_{y}=0$, the components of the dielectric tensor given by \cite%
{Alexandrov1984} reduce to
\begin{eqnarray}
\varepsilon _{xx} &=&\varepsilon _{yy}=1-\sum_{s}\left( \frac{\omega _{Ps}}{%
\omega }\right) ^{2}\frac{1}{2}\sum_{n=\pm 1}\frac{\xi _{s,0}}{\xi _{s,n}}%
J_{+}\left( \xi _{s,n}\right) ;  \nonumber \\
\varepsilon _{xy} &=&-\varepsilon _{yx}=i\sum_{s}\left( \frac{\omega _{Ps}}{%
\omega }\right) ^{2}\frac{1}{2}\sum_{n=\pm 1}n\frac{\xi _{s,0}}{\xi _{s,n}}%
J_{+}\left( \xi _{s,n}\right) ;  \nonumber \\
\varepsilon _{xz} &=&\varepsilon _{zx}=\varepsilon _{yz}=\varepsilon _{zy}=0;
\nonumber \\
\varepsilon _{zz} &=&1+\sum_{s}\left( \frac{\omega _{Ps}}{k_{z}V_{Ts}}%
\right) ^{2}\left[ 1-J_{+}\left( \xi _{s,0}\right) \right] ,  \label{e}
\end{eqnarray}%
where $\xi _{s,n}=\left( \omega -k_{z}V_{s}+n\omega _{Bs}\right) /\left(
k_{z}V_{Ts}\right) $, $\omega _{Ps}$\ ($\omega _{Bs}$) is the plasma
(cyclotron) frequency. Instead of the plasma dispersion function $W\left(
x\right) $, we use the function
\begin{equation}
J_{+}\left( x\right) =-i\sqrt{\frac{\pi }{2}}xW\left( \frac{x}{\sqrt{2}}%
\right) =x\exp \left( -\frac{x^{2}}{2}\right) \int_{i\infty }^{x}dt\exp
\left( \frac{t^{2}}{2}\right) ,  \label{j}
\end{equation}%
introduced by Alexandrov et al. (1984). It has the following asymptotic
expansions:
\begin{equation}
J_{+}\left( x\right) =x^{2}+O\left( x^{4}\right) -i\sqrt{\frac{\pi }{2}}%
x\exp \left( -\frac{x^{2}}{2}\right) ,\qquad \left\vert x\right\vert \ll 1;
\label{j<}
\end{equation}%
and
\begin{equation}
J_{+}\left( x\right) =1+\frac{1}{x^{2}}+O\left( \frac{1}{x^{4}}\right)
-i\eta \sqrt{\frac{\pi }{2}}x\exp \left( -\frac{x^{2}}{2}\right) ,\qquad
\left\vert x\right\vert \gg 1,  \label{j>}
\end{equation}%
where $\eta =0$ for $\mathrm{Im}x>0$, $\eta =1$ for $\mathrm{Im}x=0$, and $%
\eta =2$ for $\mathrm{Im}x<0$.

In the case of parallel propagation, the dispersion equation (\ref{det})
splits into two independent equations,
\begin{equation}
\varepsilon _{xx}\pm i\varepsilon _{xy}=\left( \frac{ck_{z}}{\omega }\right)
^{2},  \label{DR+-}
\end{equation}%
describing left-hand (sign -) and right-hand (sign +) polarized
electromagnetic waves. In what follows we consider the left-hand polarized
Alfv\'{e}n branch undergoing the compensated-current instability. Taking
into account quasineutrality (\ref{q=0}) and current compensation (\ref{j=0}%
), equation (\ref{DR+-}) for Alfv\'{e}n waves can be written as
\begin{equation}
\left( \frac{\omega }{\omega _{Bi}}\right) ^{2}-\frac{n_{b}}{n_{0}}%
A_{k,\omega }\frac{\omega }{\omega _{Bi}}-\left( \frac{k_{z}V_{A}}{\omega
_{Bi}}\right) ^{2}+\frac{n_{b}}{n_{0}}A_{k,\omega }\frac{k_{z}V_{b}}{\omega
_{Bi}}=0,  \label{DR}
\end{equation}%
where
\[
A_{k,\omega }=1+\frac{\omega _{Bi}}{k_{z}V_{Tb}}\frac{J_{+}\left( \xi
_{b,-1}\right) }{\xi _{b,-1}}.
\]

In the following sections we consider important limits of (\ref{DR}) typical
for the reactive CCPI instability.

\section{Dispersion relation for parallel-propagating waves}

As we are going to analyze the reactive non-resonant instability, we neglect
the contribution of the imaginary part of $J_{+}\left( \xi _{b,-1}\right) $.
Furthermore, we consider a low-frequency instability with $\left\vert \omega
/\omega _{Bi}\right\vert $ smaller than other terms in $\xi _{b,-1}$, which
allows to neglect the $\omega $-dependent part in the argument of function $%
J_{+}$. In this case (\ref{DR}) reduces to the following quadratic equation
with respect to $\omega /\omega _{Bi}$:
\begin{equation}
\left( \frac{\omega }{\omega _{Bi}}\right) ^{2}-\frac{n_{b}}{n_{0}}A_{k}%
\frac{\omega }{\omega _{Bi}}-\left( \frac{k_{z}V_{A}}{\omega _{Bi}}\right)
^{2}+\frac{n_{b}}{n_{0}}A_{k}\frac{k_{z}V_{b}}{\omega _{Bi}}=0,  \label{DR+}
\end{equation}%
where
\begin{equation}
A_{k}\equiv A_{k,0}=1+\frac{1}{k_{z}\rho _{Tb}}\frac{\mathrm{Re}J_{+}\left(
\zeta _{b}\right) }{\zeta _{b}},  \label{A0}
\end{equation}%
$\zeta _{b}=-V_{b}/V_{Tb}-1/\left( k_{z}\rho _{Tb}\right) $ and $\rho
_{Tb}=V_{Tb}/\omega _{Bi}$. To avoid misunderstanding, we stress that
although $\rho _{Tb}$ looks like the ion beam gyroradius, it is defined by
the parallel beam temperature rather than the perpendicular one and have
here a different physical meaning.

Equation (\ref{DR+}) is the second-order eigenmode equation for Alfv\'{e}n
waves modified by the ion beam and return electron current (second and
fourth terms, respectively). Its solution is straightforward:
\begin{equation}
\frac{\omega }{\omega _{Bi}}=\frac{n_{b}}{n_{0}}\frac{A_{k}}{2}+\sqrt{\left(
\frac{n_{b}}{n_{0}}\frac{A_{k}}{2}\right) ^{2}+\left( \frac{k_{z}V_{A}}{%
\omega _{Bi}}\right) ^{2}-2\frac{n_{b}}{n_{0}}\frac{A_{k}}{2}\frac{k_{z}V_{b}%
}{\omega _{Bi}}}.  \label{sol}
\end{equation}%
From (\ref{sol}) it is obvious that the instability can be driven by the
last term under the square root when $k_{z}V_{b}>0$. In what follows we
assume $V_{b}>0$ considering potentially unstable waves with $k_{z}>0$ (in
the case of $V_{b}<$ $0$, the identical instability develops for $k_{z}<$ $0$%
). In the absence of the beam, equation (\ref{sol}) reduces to the Alfv\'{e}n
wave dispersion, $\omega =k_{z}V_{A}$ at $n_{b}=0$.

The wave with dispersion (\ref{sol}) becomes unstable when the last term
under the square root dominates. This term represents effects due to the
electron current. The growth rate $\gamma =\mathrm{Im[}\omega $] of the
corresponding instability is
\begin{equation}
\frac{\gamma _{k}}{\omega _{Bi}}=\frac{V_{A}}{V_{Tb}}\sqrt{2k_{z}\rho _{Tb}%
\frac{\alpha _{b}A_{k}}{2}-\left( k_{z}\rho _{Tb}\right) ^{2}-\left( \frac{%
V_{A}}{V_{b}}\right) ^{2}\left( \frac{\alpha _{b}A_{k}}{2}\right) ^{2}}.
\label{g}
\end{equation}%
Here we introduce the cumulative destabilizing parameter
\begin{equation}
\alpha _{b}=\frac{n_{b}}{n_{0}}\frac{V_{b}}{V_{A}}\frac{V_{Tb}}{V_{A}}\equiv
\bar{j}_{b}\bar{V}_{Tb}  \label{a}
\end{equation}%
that includes all beam parameters. One can think of it as of product of the
normalized beam current $\bar{j}_{b}=n_{b}V_{b}/\left( n_{0}V_{A}\right) $
and velocity spread $\bar{V}_{Tb}=V_{Tb}/V_{A}$. The growth rate (\ref{g})
will be analyzed below analytically and numerically, and its scalings will
be found in some important limits. It is interesting to note that
the (right-hand polarized) magnetosonic instability can be obtained from
the above equation by changing the sign of the first term under the square root
(the magnetosonic instability hence requires $k_{z}V_{b}<0$).

\section{Compensated-current instability driven by hot ion beams}

Under hot beams we mean the beams with the thermal velocity spread
significantly larger than the bulk velocity, $V_{Tb}\gg $\ $V_{b}$. For such
beams, the growth rate (\ref{g}) can be simplified by neglecting the small
term $V_{b}/V_{Tb}$\ in $\xi _{b,-1}$. The argument of $J_{+}$ is then
simplified to $\xi _{b,-1}\approx $ $-1/\left( k_{z}\rho _{Tb}\right) \equiv
$ $\zeta _{b}$. In this case $\gamma _{k}$\ depends on the normalized
parallel wavenumber $k_{z}\rho _{Tb}$\ and two dimensionless bulk
parameters: $V_{A}/V_{b}$\ and $\alpha _{b}$. Then the (maximum) instability
growth rate $\gamma _{m}=$ max$_{k}\gamma $\ appears to be function of $%
\alpha _{b}$\ and $V_{A}/V_{b}$\ only, whereas the dependence on the general
multiplier $V_{A}/V_{Tb}$\ is trivial and can be excluded by the
renormalization of $\gamma _{m}$.\textbf{\ }Note that the hot-beam condition
$V_{Tb}>$\ $V_{b}$\ restricts the applicability range of the obtained below
analytical results but, in general, does not restrict the instability range
(see also Discussions).\textbf{\ }

\subsection{Instability areas in the parameter space}

Here we find the instability threshold and the instability area in the
parameter space $(\alpha _{b},V_{A}/V_{b})$. To this end, we present the
growth rate (\ref{g}) in the following useful form:
\begin{equation}
\frac{\gamma _{k}}{\omega _{Bi}}=\frac{V_{b}}{V_{Tb}}k_{z}\rho _{Tb}\sqrt{1-%
\frac{V_{A}^{2}}{V_{b}^{2}}-\left( 1-\frac{V_{A}^{2}}{V_{b}^{2}}\frac{\alpha
_{b}}{2}\frac{A_{k}}{k_{z}\rho _{Tb}}\right) ^{2}}.  \label{g0}
\end{equation}%
From (\ref{g0}), the instability condition is obtained as
\begin{equation}
1-\frac{V_{A}^{2}}{V_{b}^{2}}>\left( 1-\frac{V_{A}^{2}}{V_{b}^{2}}\frac{%
\alpha _{b}}{2}\frac{A_{k}}{k_{z}\rho _{Tb}}\right) ^{2}.  \label{cond}
\end{equation}%
Since the right-hand side of (\ref{cond}) is positive, it is obvious that
only super-Alfv\'{e}n beams, $V_{b}>V_{A}$, may trigger instability.
Therefore, the absolute threshold for the beam velocity is $V_{b}^{\mathrm{%
thr}}=V_{A}$ and the system is stable with respect to reactive CCPI for all $%
V_{b}<$ $V_{A}$.

Using (\ref{cond}), it is also possible to find the threshold for $\alpha
_{b}$ analytically. First, solving (\ref{cond}) with respect to the $k_{k}$%
-dependent term $A_{k}/\left( k_{z}\rho _{Tb}\right) $, we find that the
unstable wavenumbers $k_{z}$ should satisfy
\begin{equation}
\frac{2}{\alpha _{b}}\frac{1}{1+\sqrt{1-\frac{V_{A}^{2}}{V_{b}^{2}}}}<\frac{%
A_{k}}{k_{z}\rho _{Tb}}<\frac{2}{\alpha _{b}}\frac{V_{b}^{2}}{V_{A}^{2}}%
\left( 1+\sqrt{1-\frac{V_{A}^{2}}{V_{b}^{2}}}\right) .  \label{ab}
\end{equation}%
When the velocity threshold is exceeded, $V_{b}>V_{A}$, the right boundary
of (\ref{ab}) is always larger than the left boundary making the interval
between them non-empty. As the function $A_{k}/\left( k_{z}\rho _{Tb}\right)
$ is limited by the maximum value $\mathrm{max}_{k}\left[ A_{k}/\left(
k_{z}\rho _{Tb}\right) \right] \approx $ $0.411$ acheaved at $k_{z}^{\ast
}\rho _{Tb}\approx 1.541$, the condition (\ref{ab}) can only be satisfied
for sufficiently large $\alpha _{b}$. From the left-hand inequality, it
immediately follows the instability condition for $\alpha _{b}$ and the
corresponding threshold:
\begin{equation}
\alpha _{b}>\alpha _{b}^{\mathrm{thr}}=\frac{2}{\mathrm{max}_{k}\left[ \frac{%
A_{k}}{k_{z}\rho _{Tb}}\right] \left( 1+\sqrt{1-\frac{V_{A}^{2}}{V_{b}^{2}}}%
\right) }=\frac{4.866}{1+\sqrt{1-\left( \frac{V_{A}}{V_{b}}\right) ^{2}}}.
\label{athr}
\end{equation}

\begin{figure}[tbp]
\centering
\includegraphics[width=9cm]{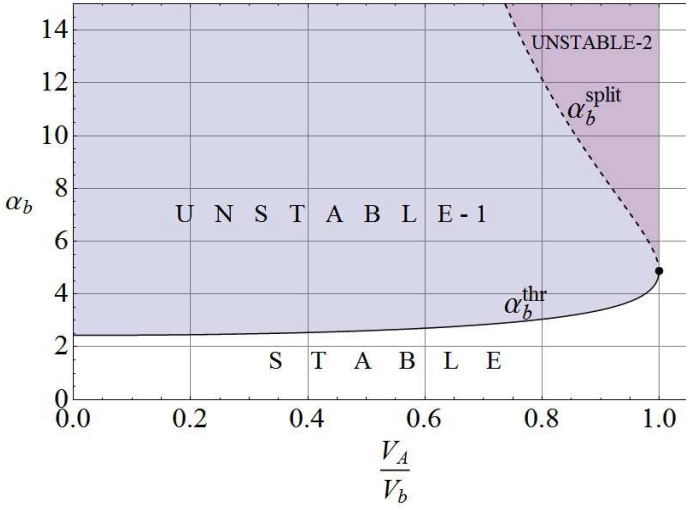}
\caption{The instability threshold $\protect\alpha _{b}^{\mathrm{thr}}$ in
the parameter space $(\protect\alpha _{b},V_{A}/V_{b})$ (solid line); the
CCPI develops at all $\protect\alpha _{b}>\protect\alpha _{b}^{\mathrm{thr}}$%
. The dashed line shows the split threshold $\protect\alpha _{b2}^{\mathrm{%
thr}}$ above which there are two separate ranges of unstable wavenumbers $%
k_{z}$. }
\label{f1}
\end{figure}

The instability condition $\alpha _{b}>\alpha _{b}^{\mathrm{thr}}$ is
satisfied above the threshold curve defined by (\ref{athr}), which is shown
in Fig. 1 by the solid line. The unstable area above this curve in the
parameter space $(\alpha _{b},V_{A}/V_{b})$ is shaded. The dependence of the
threshold $\alpha _{b}^{\mathrm{thr}}$ on $V_{A}/V_{b}$ is rather weak, it
grows from the minimal value $\alpha _{b}^{\mathrm{thr}}\approx 2.43$ at $%
V_{A}/V_{b}\rightarrow 0$ to the maximal value $\alpha _{b}^{\mathrm{thr}%
}\approx $ $4.87$ at $V_{A}/V_{b}\rightarrow 1$. The absolute threshold for $%
\alpha _{b}$ below which the system is stable is $\alpha _{b\mathrm{min}}^{%
\mathrm{thr}}\approx 2.43$. The meaning of the right boundary in (\ref{ab}),
shown in Fig. 1 by the dashed line, is clarified in the following subsection.

In terms of the normalized beam current $\bar{j}_{b}=n_{b}V_{b}/\left(
n_{0}V_{A}\right) $\ and velocity spread $\bar{V}_{Tb}=V_{Tb}/V_{A}$, (\ref%
{athr}) can be written as $\bar{j}_{b}\bar{V}_{Tb}>\alpha _{b}^{\mathrm{thr}}
$. Then the instability condition for the beam velocity spread reads as
\begin{equation}
\bar{V}_{Tb}>\bar{V}_{Tb}^{\mathrm{thr}}=\frac{\alpha _{b}^{\mathrm{thr}}}{%
\bar{j}_{b}}.  \label{V_thr}
\end{equation}%
This threshold-like condition is an important new result quantitatively
demonstrating the destabilizing effect of the beam velocity spread. It shows
the threshold above which the beam velocity spread triggers the instability
even for weak beams.

Similarly, the threshold condition for the beam current can be written as%
\textbf{\ }%
\begin{equation}
\mathbf{\ }\bar{j}_{b}>\bar{j}_{b}^{\mathrm{thr}}\approx \frac{\alpha _{b}^{%
\mathrm{thr}}}{\bar{V}_{Tb}},  \label{j_thr}
\end{equation}%
which quantifies the range of unstable beam currents. Again, it is seen that
even very weak ion beams can activate CCPI provided their velocity spreads
are sufficiently high. In particular, the beam current required for the
instability can be many orders of magnitude smaller than the Alfv\'{e}n
current.

Note that $\alpha _{b}^{\mathrm{thr}}$\ varies slowly for fast super-Alfv%
\'{e}nic beams and can be approximated as $\alpha _{b}^{\mathrm{thr}}\approx
$ $2.5$ at $V_{b}/V_{A}>$ $3$.\ For rough estimations, in all velocity range
$\alpha _{b}^{\mathrm{thr}}$\ can be replaced by its average value $3.5$.

\subsection{Unstable wavenumber ranges}

Properties of CCPI are illustrated further by Figs. 2 and 3 showing all
three terms of the condition (\ref{ab}): the left and right boundaries, and
the function $\left( k_{z}\rho _{Tb}\right) ^{-1}A_{k}$. The unstable ranges
where (\ref{ab}) is satisfied are shaded. A regular single-peak behavior of
the function $\left( k_{z}\rho _{Tb}\right) ^{-1}A_{k}$, as is seen in Figs.
2 and 3, allows us to investigate how the unstable wavenumber range evolves
with $\alpha _{b}$.

When $\alpha _{b}$ increases being still smaller than $\alpha _{b}^{\mathrm{%
thr}}$, the left boundary of (\ref{ab}) decreases remaning above the maximum
of $\left( k_{z}\rho _{Tb}\right) ^{-1}A_{k}$. In this case there are no
unstable wavenumbers and the system is stable. Once $\alpha _{b}$ rises
above $\alpha _{b}^{\mathrm{thr}}$, the decreasing left boundary of (\ref{ab}%
) drops below the maximum of $\left( k_{z}\rho _{Tb}\right) ^{-1}A_{k}$ and
the unstable wavenumber range $k_{z1}<k_{z}<k_{z2}$ appears, where $k_{z1}$
and $k_{z2}$ are lower and upper roots of equation
\begin{equation}
\frac{A_{k}}{k_{z}\rho _{Tb}}=\frac{2}{\alpha _{b}}\frac{1}{1+\sqrt{1-\frac{%
V_{A}^{2}}{V_{b}^{2}}}}.  \label{kz12}
\end{equation}

As long as $\alpha _{b}$ is not far from the threshold $\alpha _{b}^{\mathrm{%
thr}}$, there is a single unstable wavenumber interval surrounding $%
k_{z}^{\ast }\rho _{Tb}\approx $ $1.54$. This situation is illustrated in
Fig. 2, where $V_{A}/V_{b}=$ $0.9$, $\alpha _{b}^{\mathrm{thr}}\approx 3.4$,
and $\alpha _{b}=$ $6>\alpha _{b}^{\mathrm{thr}}$. However, when $\alpha
_{b} $ increases further, the right boundary of (\ref{ab}) also drops below
the maximum of $\left( k_{z}\rho _{Tb}\right) ^{-1}A_{k}$, which happens at
\begin{equation}
\alpha _{b}>\alpha _{b}^{\mathrm{split}}=4.866\frac{V_{b}^{2}}{V_{A}^{2}}%
\left( 1+\sqrt{1-\frac{V_{A}^{2}}{V_{b}^{2}}}\right) .  \label{split}
\end{equation}%
In this case, shown in Fig. 3 for $\alpha _{b}=$ $9$, the right-hand side
inequality of (\ref{ab}) is not satisfied in the range $k_{z1}^{\prime
}<k_{z}<k_{z2}^{\prime }$, where $k_{z1}^{\prime }$ and $k_{z2}^{\prime }$
are the lower and upper roots of equation
\begin{equation}
\frac{A_{k}}{k_{z}\rho _{Tb}}=\frac{2}{\alpha _{b}}\frac{V_{b}^{2}}{V_{A}^{2}%
}\left( 1+\sqrt{1-\frac{V_{A}^{2}}{V_{b}^{2}}}\right) .  \label{kz34}
\end{equation}%
Instead of unstable, we have now a prohibited wavenumber range around $%
k_{z}^{\ast }\rho _{Tb}\approx $ $1.54$. As a result, the unstable
wavenumber range splits into two: the first unstable range is $k_{z1}<$ $%
k_{z}<$ $k_{z1}^{\prime }$ and the second $k_{z2}^{\prime }<k_{z}<k_{z2}$.

The split threshold $\alpha _{b}^{\mathrm{split}}$ (\ref{split}) is shown in
Fig. 1 by the dashed line. For parameter values above this line, the
instability develops in two wavenumber ranges mentioned above. These
unstable ranges are shown in Fig. 3 by the shaded areas.

\begin{figure}[]
\centering
\includegraphics[width=9cm]{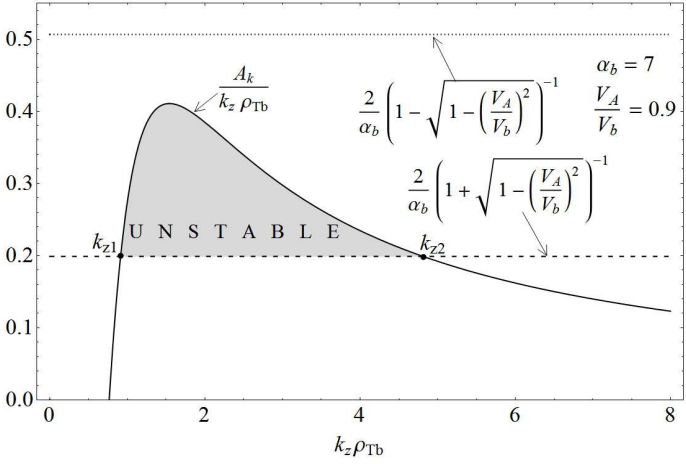}
\caption{Illustration of the condition (\protect\ref{ab}) for $V_{A}/V_{b}=$
$0.9$ and $\protect\alpha _{b}=6$. In this case $\protect\alpha _{b}^{%
\mathrm{thr}}<$ $\protect\alpha _{b}<$ $\protect\alpha _{b}^{\mathrm{split}}$
and there is only one unstable wavenumber range (shaded area). }
\label{f2}
\end{figure}

\begin{figure}[]
\centering
\includegraphics[width=9cm]{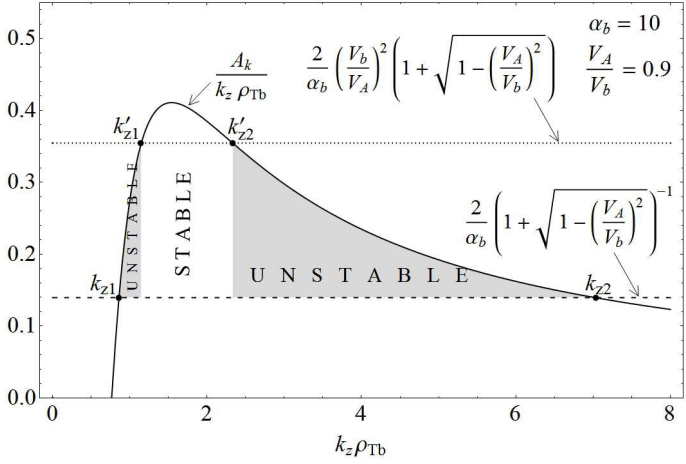}
\caption{Illustration of the condition (\protect\ref{ab}) for $V_{A}/V_{b}=$
$0.9$ and $\protect\alpha _{b}=10$. In this case $\protect\alpha _{b}>%
\protect\alpha _{b}^{\mathrm{split}}$ and there two unstable wavenumber
ranges presented by two shaded areas. }
\label{f3}
\end{figure}

Furthermore, Fig. 4 shows the $\alpha _{b}$-dependence of the unstable
wavenumber ranges, where the outer and inner boundaries are defined,
respectively, by the left-hand and right-hand margins of (\ref{ab}). It is
seen that below $\alpha _{b}^{\mathrm{thr}}$ there is no instability, at $%
\alpha _{b}^{\mathrm{thr}}<\alpha _{b}<\alpha _{b}^{\mathrm{split}}$ there
is a single unstable range of $k_{z}$, and above $\alpha _{b}^{\mathrm{split}%
}$ there are two unstable ranges.

\begin{figure}[]
\centering
\includegraphics[width=9cm]{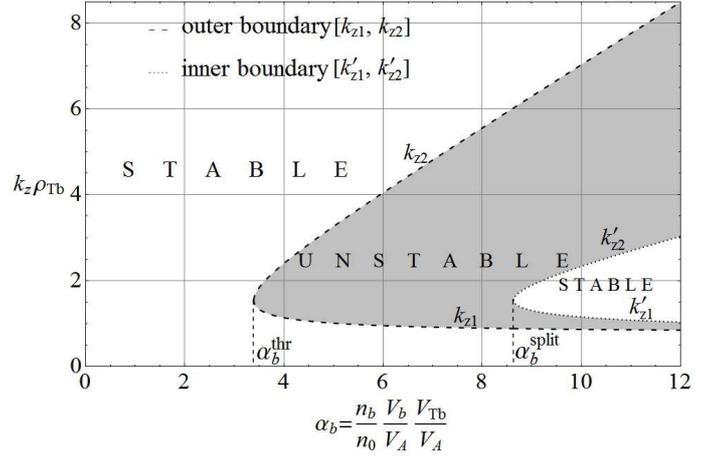}
\caption{Unstable wavenumber ranges in the $(\protect\alpha _{b},k_{z})$
plane for $V_{A}/V_{b}=$ $0.9$. The outer boundary is defined by the
left-hand side and the inner boundary by the right-hand side of the
condition (\protect\ref{ab}). It is seen that below $\protect\alpha _{b}^{%
\mathrm{thr}}$ there is no instability, at $\protect\alpha _{b}^{\mathrm{thr}%
}<$ $\protect\alpha _{b}<$ $\protect\alpha _{b}^{\mathrm{split}}$ there is a
single unstable range of $k_{z}$, and above $\protect\alpha _{b}^{\mathrm{%
split}}$ there are two unstable ranges.}
\label{f4}
\end{figure}

From Fig. 3 it is obvious that $k_{z1}\rho _{Tb}$ and $k_{z1}^{\prime }\rho
_{Tb}$ are located between $k_{z}\rho _{Tb}\approx 0.77$, where $\left(
k_{z}\rho _{Tb}\right) ^{-1}A_{k}$ is zero, and $k_{z}^{\ast }\rho
_{Tb}\approx 1.54$, where $\left( k_{z}\rho _{Tb}\right) ^{-1}A_{k}$ is
maximal. This wavenumber range corresponds to $-1.3<\zeta _{b}<-0.65$, where
$\mathrm{Re}J_{+}\left( \zeta _{b}\right) $ can be approximated by the liner
numerical fit
\begin{equation}
\mathrm{Re}J_{+}\left( \zeta _{b}\right) \approx -0.275-\zeta _{b}.
\label{Japprox}
\end{equation}%
Using this in (\ref{kz12}) and (\ref{kz34}), we find $k_{z1}$ and $%
k_{z1}^{\prime }$ as
\begin{equation}
k_{z1}\rho _{Tb}\approx \frac{1}{0.64+\sqrt{0.41-\frac{2}{\alpha _{b}}\left(
1+\sqrt{1-\frac{V_{A}^{2}}{V_{b}^{2}}}\right) ^{-1}}};  \label{k1}
\end{equation}%
\begin{equation}
k_{z1}^{\prime }\rho _{Tb}\approx \frac{1}{0.64+\sqrt{0.41-\frac{2}{\alpha
_{b}}\frac{V_{b}^{2}}{V_{A}^{2}}\left( 1+\sqrt{1-\frac{V_{A}^{2}}{V_{b}^{2}}}%
\right) }}.  \label{k11}
\end{equation}%
From these expressions we see that with increasing $\alpha _{b}$ the
difference between $k_{z1}$ and $k_{z1}^{\prime }$ decreases, $%
k_{z1}^{\prime }\rho _{Tb}\rightarrow $ $k_{z1}\rho _{Tb}\rightarrow $ $0.766
$, and the first unstable range becomes very narrow.

On the other hand, the roots $k_{z2}\rho _{Tb}$ and $k_{z2}^{\prime }\rho
_{Tb}$ bounding the second unstable range, are located above $k_{z}^{\ast
}\rho _{Tb}\approx 1.54$, where $\zeta _{b}>$ $-0.65$. Then, using the small
argument expansion (\ref{j<}) for $\mathrm{Re}J_{+}\left( \zeta _{b}\right) $%
, we find
\begin{equation}
k_{z2}^{\prime }\rho _{Tb}=\frac{\alpha _{b}}{2}\frac{V_{A}^{2}}{V_{b}^{2}}%
\left( 1+\sqrt{1-\frac{V_{A}^{2}}{V_{b}^{2}}}\right) ^{-1}-\frac{2}{\alpha
_{b}}\frac{V_{b}^{2}}{V_{A}^{2}}\left( 1+\sqrt{1-\frac{V_{A}^{2}}{V_{b}^{2}}}%
\right) ;  \label{k22}
\end{equation}%
\begin{equation}
k_{z2}\rho _{Tb}=\frac{\alpha _{b}}{2}\left( 1+\sqrt{1-\frac{V_{A}^{2}}{%
V_{b}^{2}}}\right) -\frac{2}{\alpha _{b}}\left( 1+\sqrt{1-\frac{V_{A}^{2}}{%
V_{b}^{2}}}\right) ^{-1}.  \label{k2}
\end{equation}%
At large $\alpha _{b}$, both the width of the second unstable range $%
k_{z2}\rho _{Tb}-k_{z2}^{\prime }\rho _{Tb}$ and the gap between the
unstable ranges $k_{z2}^{\prime }\rho _{Tb}-k_{z1}^{\prime }\rho _{Tb}$ grow
linearly with $\alpha _{b}$.

Summarizing above, the most important analytical result obtained here is the
instability boundary $\alpha _{b}^{\mathrm{thr}}$ in the parameter space ($%
V_{A}/V_{b};\alpha _{b}$), which can be used directly to analyze
observational data. The compensated-current systems with $V_{A}/V_{b}<1$ and
$\alpha _{b}>\alpha _{b}^{\mathrm{thr}}$ are unstable. The unstable area in
the parameter space ($V_{A}/V_{b};\alpha _{b}$) is divided further by $%
\alpha _{b}^{\mathrm{split}}$ into two unstable sub-areas: $\alpha _{b}^{%
\mathrm{thr}}<\alpha _{b}<\alpha _{b}^{\mathrm{split}}$ with one unstable
wavenumber range, and $\alpha _{b}>\alpha _{b}^{\mathrm{split}}$ with two
unstable wavenumber ranges.

\subsection{Instability growth rate}

Once $\alpha _{b}$ rises above $\alpha _{b}^{\mathrm{thr}}$, an unstable
range between $k_{z1}$ and $k_{z2}$ appear. The instability growth rate (\ref%
{g}) as function of $k_{z}$ is shown in Fig. 5. The plasma parameters $%
\alpha _{b}$ and $V_{A}/V_{b}$ in this figure are chosen in such a way as to
illustrate behavior of CCPI in the unstable wavenumber ranges found above.
So, the case $\alpha _{b}=$ $6$ with one unstable wavenumber range is shown
by the dashed line and the case $\alpha _{b}=$ $10$ with two unstable
wavenumber ranges is shown by the solid lines. The dotted curve in Fig. 5 is
for the case $\alpha _{b}=$ $8$ that is close to the splitting threshold. It
is seen that when the right instability boundary in (\ref{ab}) approaches
the maximum of function $\left( k_{z}\rho _{Tb}\right) ^{-1}A_{k}$, the
valley and the second peak in $\gamma _{k}$ appear. This happens at $\alpha
_{b}>$ $\alpha _{b}^{\mathrm{pl}}$, where
\begin{equation}
\alpha _{b}^{\mathrm{pl}}\approx \frac{V_{b}^{2}}{V_{A}^{2}}\left( 4.1+\sqrt{%
16-15\frac{V_{A}^{2}}{V_{b}^{2}}}\right)   \label{apl}
\end{equation}%
is the value of $\alpha _{b}$ at which a local "plateau" in $\gamma _{k}$
occurs at the wavenumber where $\partial \gamma _{k}/\partial k_{z}=$ $0$
and $\partial ^{2}\gamma _{k}/\partial k_{z}^{2}=$ $0$. For all $\alpha
_{b}>\alpha _{b}^{\mathrm{pl}}$, the secondary peak of $\gamma _{k}$ exists
at $k_{z}<k_{z}^{\ast }$. Since $\alpha _{b}^{\mathrm{pl}}<\alpha _{b}^{%
\mathrm{split}}$, the secondary peak arises before the interval of
prohibited wavenumbers $k_{z1}^{\prime }<k_{z}<k_{z2}^{\prime }$ appears.

It is seen that CCPI is stronger and the most unstable wavenumbers are
larger for larger $\alpha _{b}$. The secondary peak that appears at $\alpha
_{b}>\alpha _{b}^{\mathrm{pl}}$ is lower than the main peak. These trends
are confirmed below analytically.

The most unstable wavenumber and the corresponding maximum growth rate $%
\gamma _{\mathrm{max}}$ can be found by maximizing (\ref{g}) with respect to
$k_{z}$, $\gamma _{\mathrm{max}}=$ $\mathrm{max}_{k}\left( \gamma
_{k}\right) $, which we call the CCPI growth rate.

\begin{figure}[]
\centering
\includegraphics[width=9cm]{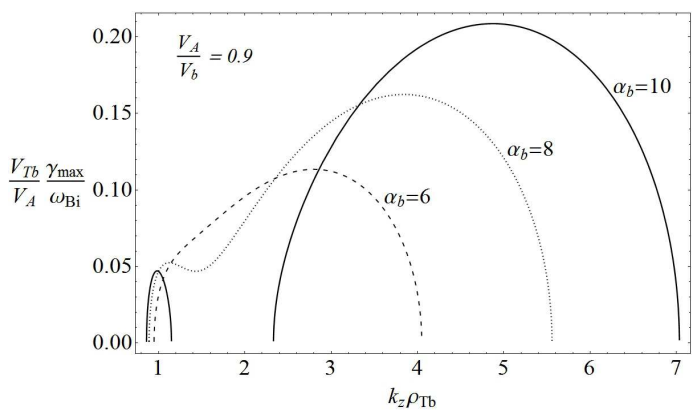}
\caption{Wavenumber dependence of the instability growth rate driven by
super-Alfv\'{e}nic ion beams with $V_{A}/V_{b}=$ $0.9$ for three values of $%
\protect\alpha _{b}$: $\protect\alpha _{b}=6,$ $8$, and $10$. For larger $%
\protect\alpha _{b}$, the unstable area and the maximum growth rate extend
to larger $k_{z}\protect\rho _{Tb}$. }
\label{f5}
\end{figure}

The normalized CCPI growth rate $\gamma _{\mathrm{max}}/\omega _{Bi}$ as
function of $n_{b}/n_{0}$ and $V_{b}/V_{A}$ is shown in Fig. 6 for hot beam
with $V_{Tb}/V_{A}=$ $10^{2}$. It is seen that $\gamma _{\mathrm{max}}$
increases fast, roughly proportional to both $n_{b}/n_{0}$ and $V_{b}/V_{A}$%
, which means it is proportional to the current $~n_{b}V_{b}$. This behavior
agrees with the current nature of CCPI confirmed below analytically by (\ref%
{11c})-(\ref{11d}).

The threshold for $n_{b}/n_{0}$ ($V_{b}/V_{A}$) is lower for smaller $%
V_{b}/V_{A}$ ($n_{b}/n_{0}$), in agreement with (\ref{athr}). In particular,
the velocity threshold $V_{b}^{\mathrm{thr}}/V_{A}$ decreases with $%
n_{b}/n_{0}$ and reaches the minimal value $V_{b}^{\mathrm{thr}%
}/V_{A}\rightarrow 1$ when $n_{b}V_{Tb}/\left( n_{0}V_{A}\right) \rightarrow
1$.

\begin{figure}[]
\centering
\includegraphics[width=9cm]{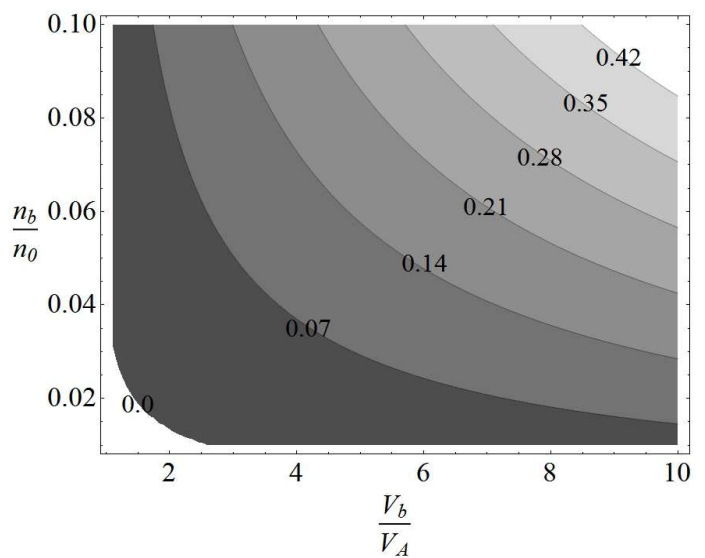}
\caption{Normalized growth rate $\protect\gamma _{\mathrm{max}}/\protect%
\omega _{Bi}$ as function of $n_{b}/n_{0}$ and $V_{b}/V_{A}$ for hot beam
with $V_{Tb}/V_{A}=$ $10^{2}$. $\protect\gamma _{\mathrm{max}}$ is regularly
increasing with both $n_{b}/n_{0}$ and $V_{b}/V_{A}$ once the threshold is
exceeded. }
\label{f6}
\end{figure}

\begin{figure}[]
\centering\includegraphics[width=9cm]{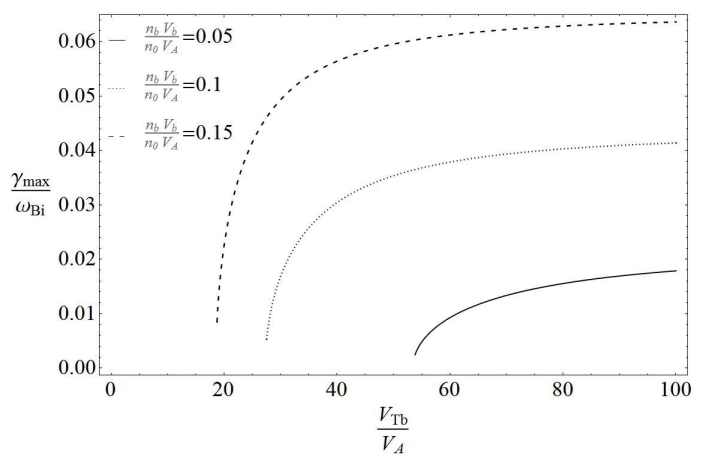}
\caption{Normalized growth rate $\protect\gamma _{\mathrm{max}}/\protect%
\omega _{Bi}$ as function of $V_{Tb}/V_{A}$ for $n_{b}V_{b}/\left(
n_{0}V_{A}\right) =$ $0.05$ (solid curve), $0.1$ (dotted curve), and $0.15$
(dashed curve). $\protect\gamma _{\mathrm{max}}$ is regularly growing with $%
V_{Tb}$ but this grows is quickly saturated. Larger currents $%
n_{b}V_{b}/\left( n_{0}V_{A}\right) $ result in larger $\protect\gamma _{%
\mathrm{max}}$ for all $V_{Tb}$. }
\label{f7}
\end{figure}

Dependence of $\gamma _{\mathrm{max}}$ on the thermal velocity $V_{Tb}$ is
somehow different (see Fig. 7). First, near the threshold, $\gamma _{\mathrm{%
max}}$ grows very fast with $V_{Tb}$, but then its growth is quickly
saturated. Already at $V_{Tb}\gtrsim $ $2V_{Tb}^{\mathrm{thr}}$, $\gamma _{%
\mathrm{max}}$ becomes virtually independent of $V_{Tb}$.

To understand this behavior, we proceed with the analytical analysis. Here
we take into account the fact that in the wavenumber range $k_{z}\rho
_{Tb}>k_{z}^{\ast }\rho _{Tb}\approx 1.54$, where the growth rate attains
its maximum, the low-$\zeta _{b}$ approximation $\mathrm{Re}J_{+}\left(
\zeta _{b}\right) \approx $ $-\left( k_{z}\rho _{Tb}\right) ^{-2}$ is valid.
Thus, using $A_{k}=1+\mathrm{Re}J_{+}\left( \zeta _{b}\right) \approx $ $%
1-\left( k_{z}\rho _{Tb}\right) ^{-2}$ in (\ref{g0}), we find the the
following approximation for the maximum of $\gamma _{k}$:
\begin{equation}
\frac{\gamma _{\mathrm{max}}}{\omega _{Bi}}\approx \frac{1}{2}\frac{%
n_{b}V_{b}}{n_{0}V_{A}}\sqrt{\left( 1-\frac{V_{A}^{2}}{V_{b}^{2}}\right) %
\left[ 1-\left( \frac{\alpha _{b}^{\mathrm{thr}}}{\alpha _{b}}\right) ^{2}%
\right] }.  \label{gm}
\end{equation}%
This maximum occurs at
\begin{equation}
k_{z}^{\mathrm{m}}\rho _{Tb}\approx \frac{\alpha _{b}}{2}+\frac{V_{b}}{V_{Tb}%
}\frac{2}{\alpha _{b}}\left( 1-2\frac{V_{A}^{2}}{V_{b}^{2}}\right) .
\label{km}
\end{equation}%
The last term in the square parentheses in (\ref{gm}) is adjusted by
replacing the approximate numerical value $\sqrt{8}$ by $\alpha _{b}^{%
\mathrm{thr}}$ to make it compatible with the exact $\alpha _{b}$-threshold (%
\ref{athr}). We verified numerically that the approximation (\ref{gm}) is
good for arbitrary $\alpha _{b}$, both near the threshold and far from it.
In general, with the larger beam velocity and/or temperature, the smaller
beam density is needed for instability.

The explicit dependence of the instability growth rate on the beam velocity
spread $\bar{V}_{Tb}$\ follows from (\ref{gm}):\textbf{\ }%
\begin{equation}
\frac{\gamma _{\mathrm{max}}}{\omega _{Bi}}\approx \frac{\bar{j}_{b}}{2}%
\sqrt{\left( 1-\frac{1}{\bar{V}_{b}^{2}}\right) \left[ 1-\left( \frac{\alpha
_{b}^{\mathrm{thr}}}{\bar{j}_{b}\bar{V}_{Tb}}\right) ^{2}\right] }.
\label{gT}
\end{equation}%
\textbf{\ }It is seen that $\gamma _{\mathrm{max}}$\ increases quickly with $%
\bar{V}_{Tb}$ once the threshold is overcomed, $\bar{V}_{Tb}\gtrsim $ $%
\alpha _{b}^{\mathrm{thr}}/\bar{j}_{b}$. The fast increase of $\gamma _{%
\mathrm{max}}$\ reflects the instability response to the progressive
demagnetization of the beam ions as their velocity spread increases above
the threshold.

However, when $\bar{V}_{Tb}$\ becomes large enough, $\bar{V}_{Tb}\gtrsim $ $%
3\alpha _{b}^{\mathrm{thr}}/\bar{j}_{b}$, the term containing it becomes
negligibly small and $\gamma _{\mathrm{max}}$\ becomes virtually independent
of $\bar{V}_{Tb}$. In this high-temperature regime the beam ions are fully
demagnetized and the further increase of $\bar{V}_{Tb}$ does not affect the
instability any more. This regime corresponds to the well over-threshold
limit $\left( \alpha _{b}^{\mathrm{thr}}/\alpha _{b}\right) ^{2}\ll $ $1$
where $\gamma _{\mathrm{max}}$ simplifies to
\begin{equation}
\frac{\gamma _{\mathrm{max}}}{\omega _{Bi}}=\frac{1}{2}\frac{n_{b}V_{b}}{%
n_{0}V_{A}}\sqrt{1-\frac{V_{A}^{2}}{V_{b}^{2}}}.  \label{11c}
\end{equation}%
The familiar threshold velocity of the beam, $V_{b}^{\mathrm{thr}}=V_{A}$,
is still present in (\ref{11c}), but the temperature dependence is already
missed, as can be observed in Fig. 7 at large $V_{Tb}$.

The maximum growth rate (\ref{11c}) simplifies further for the fast beams
with $V_{b}/V_{A}>$ $3$,
\begin{equation}
\frac{\gamma _{\mathrm{max}}}{\omega _{Bi}}\approx \frac{1}{2}\frac{%
n_{b}V_{b}}{n_{0}V_{A}},  \label{11d}
\end{equation}%
with the most unstable parallel wavenumber $k_{z}^{\mathrm{m}}\rho
_{Tb}=\alpha _{b}/2$. The asymptotic scaling (\ref{11d}) recovers the
scaling obtained by Bell (2004). As is seen from Fig. 7, expressions (\ref%
{11c}) and (\ref{11d}) provide good estimations for $\gamma _{\mathrm{max}}$
at $V_{Tb}>$ $2V_{Tb}^{\mathrm{thr}}$, which also quantifies the meaning of
"asymptotic regime" in terms of $V_{Tb}$. It appears that the expressions
found by Bell are only valid in this asymptotic regime.

For $\alpha _{b}>$ $\alpha _{b}^{\mathrm{pl}}$, the secondary peak arises at
$k_{z}\rho _{Tb}<1.54$, where we can use approximation (\ref{Japprox}). Then
for this peak we obtain the local maximum
\begin{equation}
\frac{\gamma _{\mathrm{m2}}}{\omega _{Bi}}\approx \frac{V_{A}}{V_{Tb}}k_{z}^{%
\mathrm{m2}}\rho _{Tb}\sqrt{\left( 1-\frac{V_{A}^{2}}{V_{b}^{2}}\right) }
\label{gm2}
\end{equation}%
attained at
\begin{equation}
k_{z}^{\mathrm{m2}}\rho _{Tb}\approx 0.765+\frac{2}{\alpha _{b}^{\ast }},
\label{km2}
\end{equation}%
where $\alpha _{b}^{\ast }=\left( \frac{V_{A}}{V_{b}}\right) ^{2}\alpha _{b}$
and we took into account that $\alpha _{b}>$ $\alpha _{b}^{\mathrm{pl}}$.
The ratio of this peak to the main peak is
\begin{equation}
\frac{\gamma _{\mathrm{m2}}}{\gamma _{\mathrm{max}}}=\frac{k_{z}^{\mathrm{m2}%
}\rho _{Tb}}{\frac{1}{2}\alpha _{b}}=\left( \frac{V_{A}}{V_{b}}\right)
^{2}\left( 0.765+\frac{2}{\alpha _{b}^{\ast }}\right) \frac{2}{\alpha
_{b}^{\ast }}.  \label{g21}
\end{equation}%
Taking into account that $\alpha _{b}^{\ast }>$ \textrm{min}$\left[ \alpha
_{b}^{\ast }\right] \approx $ $5.1$ (at $V_{A}/V_{b}\rightarrow 1$), we see
that the peak $\gamma _{\mathrm{m2}}$ is always significantly smaller than
the main peak $\gamma _{\mathrm{max}}$. The maximum ratio $\gamma _{\mathrm{%
m2}}/\gamma _{\mathrm{max}}\approx $ $0.45$ is achieved at $\alpha
_{b}\gtrsim $ $\alpha _{b}^{\mathrm{pl}}$ and $V_{A}/V_{b}\lesssim 1$.

Note that the unstable fluctuations have also a small oscillatory part Re[$%
\omega $]= $0.5\left( n_{b}/n_{0}\right) \omega _{Bi}$. For most unstable
wavenumber $k_{z}^{\mathrm{m}}\rho _{Tb}\sim \alpha _{b}/2$, the real
frequency $\omega ^{\mathrm{m}}=$ Re$\left[ \omega \right] \sim $ $k_{z}^{%
\mathrm{m}}V_{A}\left( V_{A}/V_{b}\right) $ is smaller than the frequency of
the normal Alfv\'{e}n mode $k_{z}^{\mathrm{m}}V_{A}$. Since $\gamma _{%
\mathrm{max}}\sim $ $k_{z}^{\mathrm{m}}V_{A}>$ $\omega ^{\mathrm{m}}$, the
instability is aperiodic.

\section{Parallel Alfv\'{e}n instability in particular compensated current
systems}

Let us consider two feasible applications of CCPI. First we apply our
results to the solar wind upstream of the quasi-parallel terrestrial shock,
where the plasma conditions are relatively well documented. Then we extend
the analysis to the interstellar medium around supernova remnants, assuming
the similar scalings of the beam parameters as in the terrestrial foreshock.

\subsection{Quasi-parallel terrestrial foreshock}

Hot ion beams with $V_{Tb}>$ $V_{b}>$ $V_{A}$ are regularly observed in the
solar wind upstream of the terrestrial bow shock where the shock normal is
quasi-parallel to the interplanetary magnetic field $\mathbf{B}_{0}$ %
\citep[][]{Paschmann1981,Tsurutani1981}. This ordering of characteristic
velocities suggests that CCPI driven by hot ion beams can develop in the
quasi-parallel foreshocks.

More specifically, we will use the following scalings for characteristic
beam velocities: $V_{b}\lesssim $ $V_{\mathrm{shock}}$; $V_{Tb}\sim $ $3V_{%
\mathrm{shock}}$, where the shock velocity is equal to the solar wind speed,
$V_{\mathrm{shock}}=V_{SW}$. These scalings are compatible with observations
reported by \cite{Paschmann1981} and \cite{Tsurutani1981}. Yet another beam
parameter, number density $n_{b}$, does not vary much around $n_{b}=$ $0.1$
cm$^{-3}$ \citep[][]{Paschmann1981}. In terms of the background solar-wind
density $n_{0}\sim 5-10$ cm$^{-3}$, this gives $n_{b}/n_{0}\sim 0.01-0.02$.
Taking the typical value of Alfv\'{e}n velocity, $V_{A}\approx 0.1V_{SW}$,
we obtain the cumulative destabilizing parameter $\alpha _{b}\sim 2.5-5$,
which is slightly over-threshold depending on the particular value of $V_{b}$%
. Such proximity of the system to the CCPI threshold can be a signature of
CCPI operating in the foreshock and relaxing the beam parameters towards the
threshold.

On the other hand, as is seen from Fig. 6, even slight deviations of $\alpha
_{b}$ from the threshold can make CCPI strong. So, for $V_{A}/V_{b}\sim 0.1$
and $\alpha _{b}=6$ the maximum growth rate is already high, $\gamma _{%
\mathrm{max}}\approx 0.07\omega _{Bi}$, with the most unstable wavenumbers $%
k_{z\mathrm{m}}\rho _{Tb}\gtrsim $ $2$. \cite{Narita2006} and \cite%
{Hobara2007} analyzed properties of electromagnetic fluctuations observed
around terrestrial bow shock. Most straightforwardly, our results can be
compared with the wavenumber distribution of the fluctuations in the
quasi-parallel foreshocks shown in Figure 9 by \cite{Narita2006}, where the
measured wavenumbers are normalized by the ion gyroradius. In terms of the
background ion gyroradius $\rho _{Ti}$, with the typical temperature of the
diffuse ions $T_{b}/T_{i}=4\times 10^{2}$, our most unstable wavenumbers $%
k_{z\mathrm{m}}\rho _{Ti}\sim $ $k_{z\mathrm{m}}\rho _{Tb}/20\sim $ $0.1$
map upon the major peak observed at $k_{z}\rho _{Ti}=$ $0.1$ \citep[see
upper panel in Figure 9 by ][]{Narita2006}.

In the quasi-parallel foreshock region, Narita et al. observed also another,
subdominant peak at $k_{z}\rho _{Ti}=$ $0.6$. To explain this peak by CCPI
one needs significantly lower beam temperature, $T_{b}/T_{i}\sim 10$, which
is more typical for quasi-perpendicular foreshocks. One can speculate that
CCPI can also generate this second peak. First the CCP instability develops
in the quasi-perpendicular foreshock region where the beams have required
temperatures $T_{b}/T_{i}\sim 10$, which is supported by the observed
enhancement at $k_{z}\rho _{Ti}\approx $ $0.4$. Then the unstable
fluctuations are convected in the quasi-parallel foreshock region where
their observed wavenumbers are $k_{z}\rho _{Ti}\approx $ $0.6$.

The above estimations suggest that CCPI can contribute to electromagnetic
fluctuations observed in the quasi-parallel terrestrial foreshock and impose
limitations on the parameters of the beams formed by reflected ions. Further
direct confrontations of observed values of $\alpha _{b}$\ with the
stability diagram Fig. 1 are needed to clarify the role of CCPI in the
regulation of ion-beam parameters in the foreshock.

\subsection{Foreshock regions around supernova remnants}

Supernova remnants expanding in the interstellar medium develop bow shocks
at their boundaries. These shocks propagate with high velocities $V_{\mathrm{%
shock}}\sim 2\times 10^{9}$ cm s$^{-1}$ providing a feasible source of
energy for the cosmic rays acceleration, and also for the magnetic fields
amplification. By analogy with the terrestrial bow shock, we assume that the
reflected ions occur also in the supernova foreshocks setting up a
compensated-current system. CCPI can develop in supernova foreshocks if
parameters of reflected ions (subscript $b$) satisfy $\alpha _{b}>\alpha
_{b}^{\mathrm{thr}}$, defined by (\ref{athr}).

For reasonable background density $n_{0}=10^{-2}-1$ cm$^{-3}$ and magnetic
field $B_{0}\sim 10^{-7}-10^{-5}$ G \citep{Zweibel2010}, the Alfv\'{e}n
velocity varies in the range $V_{A}=2\times 10^{4}-2\times 10^{7}$ $\mathrm{%
cm/s}$. Then the resulting Alfv\'{e}n Mach number in supernova remnants $%
M_{A}=$ $V_{\mathrm{shock}}/V_{A}=$ $10^{2}-10^{5}$ is much larger than in
Earth's bow shock. For the similar scalings as in the terrestrial
foreshocks, $n_{b}/n_{0}\sim 0.01$, $V_{b}\sim $ $0.5V_{\mathrm{shock}}$,
and $V_{Tb}\sim $ $2V_{\mathrm{shock}}$, even with the most unfavorable $V_{%
\mathrm{shock}}/V_{A}=$ $10^{2}$ the destabilizing parameter $\alpha
_{b}\sim $ $10^{2}$ is much larger than the threshold $\alpha _{b}^{\mathrm{%
thr}}\sim $ $5$. In this well over-threshold state, the CCPI operates in the
asymptotic regime (\ref{11d}) with very high growth rate $\gamma _{\mathrm{%
max}}/\omega _{Bi}\sim 0.5$. Note that this value is already at the edge of
applicability of our low-frequency approximation. Such a high growth rate
suggests that the instability modifies the beam parameters strongly, in
particular reducing the beam velocity towards the local Alfv\'{e}n velocity,
$V_{b}\gtrsim V_{A}$.

Let us compare the instability driven by the reflected ions with the similar
instability driven by cosmic rays around supernova remnants %
\citep[][]{Bell2004,Zweibel2010}. Taking the background magnetic field $%
B_{0}\gtrsim 10^{-6}$ G and the cosmic-rays flux $n_{\mathrm{CR}}V_{b}\sim
10^{4}$ cm$^{-2}$ s$^{-1}$ \citep[][]{Zweibel2010}, we estimate the
normalized current $\bar{j}_{b}^{\mathrm{CR}}\sim $ $0.026$ and the
corresponding growth rate $\gamma _{\mathrm{max}}^{\mathrm{CR}}/\omega
_{Bi}= $ $\bar{j}_{b}^{\mathrm{CR}}/2\sim $ $0.01$ around supernova
remnants. With $\omega _{Bi}\simeq 0.03$ s$^{-1}$, we get $\gamma _{\mathrm{%
max}}^{\mathrm{CR}}\simeq 2.2\times 10^{-4}$ s$^{-1}$ in absolute numbers.

The above estimations show that the CCPI instability driven by reflected
ions is much stronger than the instability driven by cosmic rays. Therefore,
the former instability can be more efficient amplifier for magnetic fields
around supernova remnants. On the other hand, a fraction of the beam ions
can be scattered back to the shock by electromagnetic fluctuations generated
by CCPI, thus providing a seed population for the further Fermi acceleration
to high cosmic-ray energies.

\section{Discussions}

A number of competing electrostatic and electromagnetic instabilities may
arise when different plasma species move with respect to each other %
\citep[see][and references therein]{Gary2005,Bret2009}. The hierarchical
structure of these instabilities depends on many parameters and remains an
open question \citep[see further discussions
by][]{Bret2010,Brown2013,Marcowith2016}.

In our setting with hot ion beams, the fast two-stream/Buneman instabilities
are quenched by the large thermal velocities that are larger than the
streaming velocities. Inspection of Fig. 3.20 by \cite{Gary2005} shows that
the thresholds of electrostatic ion-acoustic and ion-cyclotron instabilities
are significantly larger than the Alfv\'{e}nic threshold for $%
V_{Ti}/V_{A}\sim T_{e}/T_{i}\sim 1$\ typical in the terrestrial foreshock.
Among them, the electron/ion cyclotron instability has the lowest threshold
velocity which is still very high, $V_{b}^{IC}>10^{2}V_{A}$ for $%
n_{b}<0.1n_{e}$. The ion/ion acoustic instability is suppressed further by
large beam temperatures, as is seen from Fig. 3.15 by \cite{Gary2005}.
Therefore, these high-frequency electrostatic instabilities cannot compete
with CCPI in the wide range of beam velocities $1<$\ $V_{b}/V_{A}<$\ $10^{2}$%
. At larger beam velocities, $V_{b}/V_{A}>$\ $10^{2}$, the ion-acoustic and
ion-cyclotron harmonic waves can be generated by the electron-ion relative
motion. However, even in this velocity range CCPI can develop independently
as long as the mean parameters reside in the unstable area (Fig. 1), whereas
the kinetic instabilities are quickly saturated by the local quasilinear
plateaus.

Parallel-propagating left-hand and right-hand polarized instabilities have
been studied by \cite{Gary1984} and \cite{Gary1985}. Using numerical
solutions of the dispersion equation, it has been observed that the
(left-hand polarized) Alfv\'{e}nic instability becomes competitive or even
dominant when the beam ions are sufficiently hot \citep[see Fig. 8
by][]{Gary1984}. The condition $\left\vert \xi _{b,-1}\right\vert <1$\ was
used by Gary et al. to categorize this instability as ion-beam resonant,
i.e. resulting from the direct resonant coupling of the unstable mode with
the beam ions. However, kinetic and reactive effects have not been
distinguished for this mode, which did not allow to realize that above the
threshold (\ref{athr}) the instability transforms from kinetic resonant to
reactive non-resonant (see Fig. 8 and related discussions below). In the
reactive regime, the meaning of the condition $\left\vert \xi
_{b,-1}\right\vert \approx $ $\left\vert k_{z}\rho _{Tb}\right\vert ^{-1}<1$%
\ is reversed: here it indicates that the unstable perturbations become
small-scale enough to decouple from the beam ions by the demagnetization
effect. The resulting Alfv\'{e}n instability is then driven not by the
resonant interactions with the beam ions but by the bulk return current of
the magnetized electrons. The current nature of this instability is similar
to the nature of related current instability \citep[][]{Malovichko1992} that
can develop in the absence of any beams.

Interplay of the reactive and resonant left-hand Alfv\'{e}nic instabilities
also needs further investigations. Our preliminary estimations indicate that
the relative importance of the reactive destabilizing effects increases fast
once $\alpha _{b}$ rises above the threshold $\alpha _{b}^{\mathrm{thr}}$.
In Fig. 8. we show the contribution of the reactive CCPI to the total growth
rate for reference plasma parameters that may occur in foreshocks: $%
V_{Ti}/V_{A}=T_{e}/T_{i}=1$, $V_{b}/V_{A}=10$, $V_{Tb}/V_{A}=25$, and $%
n_{b}/n_{0}=0.02$.\ The corresponding total growth rate in Fig. 8 is given
by equation (\ref{g}) with the imaginary part of $J_{+}$\ taken into
account. It therefore includes both the reactive effects due to the bulk
currents and the resonant wave-particle interactions. It is seen that the
destabilizing reactive response becomes stronger than the resonant wave
response when $\alpha _{b}$\ is still not far from the threshold $\alpha
_{b}^{\mathrm{thr}}$\ ($\alpha _{b}=5\approx 2\alpha _{b}^{\mathrm{thr}}$ in
Fig. 8). The instability is thus driven mainly by the reactive effects and
can be analyzed ignoring kinetic resonant effects, as we did in the present
study. The same approach can also be applied in the immediate vicinity of
the reactive threshold if the quasilinear plateaus or other local
deformations of the velocity distributions weaken destabilizing kinetic
effects. Analytical treatment becomes more tangled when reactive and kinetic
effects are of comparable efficiency and have to be accounted
simultaneously, in which case evolution of the system becomes more complex %
\citep[cf.][]{Yoon2017}.

\begin{figure}[tbp]
\centering\includegraphics[width=9cm]{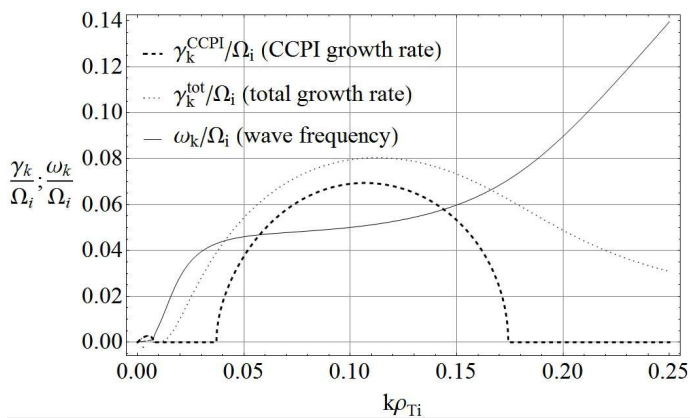}
\caption{Contribution of the reactive CCPI growth rate (dashed curve) to the
total growth rate (dotted curve) for $V_{Ti}/V_{A}=T_{e}/T_{i}=1$, $%
V_{b}/V_{A}=10$, $V_{Tb}/V_{A}=25$, and $n_{b}/n_{0}=0.02$\.{. }It is seen
that the reactive destabilizing effects dominate the instability growth rate
for this set of parameters. The wave frequency is shown by the solid line.}
\label{f8}
\end{figure}

There are also left- and right-hand polarized instabilities driven by cold
ion beams in the ion-cyclotron frequency range \citep[][]{Mecheri2007}.
These instabilities are strong when the beam velocity spread is so small
that all the beam particles (and hence the beam as a whole) are resonant. In
our settings with hot ion beams these instabilities are quenched similarly
to the two-stream/Buneman instabilities.

In the considered case of hot ion beams, $V_{b}/V_{Tb}<1$, the analytical
treatment of wavenumbers $k_{z}\rho _{Tb}<V_{Tb}/V_{b}$\ is simplified by
neglecting the term $\sim V_{b}/V_{Tb}$\ in $\xi _{b,-1}$. As the most
unstable wavenumber scales as $k_{z}\rho _{Tb}\approx $\ $\alpha _{b}/2$ (%
\ref{km}), this restriction is not stringent,\textbf{\ }%
\begin{equation}
\frac{n_{b}}{n_{0}}\left( \frac{V_{b}}{V_{A}}\right) ^{2}<2.  \label{nonFH}
\end{equation}%
This condition is opposite to the firehose instability condition \citep[see
Eq. 14 by ][]{Malovichko2014}, which means that the CCPI can operate in a
wide range of parameters below the firehose threshold. For cooler beams,
where the condition $V_{b}/V_{Tb}<1$ is violated (as, for example, in the
quasi-perpendicular foreshock regions), the analysis should be extended by
accounting for corresponding terms.

CCPI can also affect other processes in space. For example, it can limit the
field-aligned currents generated by Alfv\'{e}n-wave fluxes in the inner
magnetosphere and plasma sheet boundary layer \citep[][]{Artemyev2016}. In
the solar wind, CCPI can contribute to the regulation of relative motion of
different plasma species. It was found that many states of beaming
structures in the solar wind are close to the thresholds of magnetosonic and
Alfv\'{e}n instabilities \citep[][]{Marsch1987,Gary2000} and firehose
instability \citep[][]{Chen2016}. Since CCPI can operate close to these
thresholds (and sometimes below them), a refined analysis is needed to
decide its role in the solar wind as compared to the magnetosonic and
firehose instabilities. These are other subjects for future studies.

\section{Conclusions}

We investigated reactive non-resonant compensated-current parallel
instability (CCPI) of left-hand polarized Alfv\'{e}n waves in
compensated-current systems established by hot diluted ion beams. Ion-beam
demagnetization due to finite $k_{z}\rho _{Tb}$ is crucial for CCPI ($\rho
_{Tb}$ is based on the parallel beam temperature and hence does not
represent the beam ion gyroradius). New analytical expressions for the
instability growth rate (\ref{gm}) and threshold (\ref{athr}) are found and
analyzed.

Most important new properties of CCPI can be summarized as follows:

1. Reactive non-resonant CCPI depends on all bulk parameters of the beam:
beam density $n_{b}$, velocity $V_{b}$, and velocity spread $V_{Tb}$. All
these parameters increase the instability growth rate and can be combined in
the single destabilizing parameter $\alpha _{b}=$ $\left( n_{b}/n_{0}\right)
\left( V_{b}/V_{A}\right) \left( V_{Tb}/V_{A}\right) $. The instability
develops at $\alpha _{b}>$\textbf{\ }$\alpha _{b}^{\mathrm{thr}}$, where the
instability threshold\ \textbf{(\ref{athr})\ }varies from $\alpha _{b}^{%
\mathrm{thr}}=$ $2.43$ at $V_{A}/V_{b}\rightarrow 0$ to $\alpha _{b}^{%
\mathrm{thr}}=$ $4.87$ at $V_{A}/V_{b}\rightarrow 1$. The analytical
threshold\ \textbf{(\ref{athr})} can be directly compared with satellite
data to analyze stability of beam-plasma systems in space.

2. CCPI is strongly affected by the beam velocity spread $V_{Tb}$.\ It
defines the range of unstable beam currents, $\bar{j}_{b}\geq $ $\bar{j}%
_{b}^{\mathrm{thr}}$, with the current threshold varying in the range $\bar{j%
}_{b}^{\mathrm{thr}}=$ $\left( 2.4-4.9\right) /\bar{V}_{Tb}$.

3. The instability growth rate $\gamma _{\mathrm{max}}$ (\ref{gT})\
increases sharply with $V_{Tb}$ once the threshold $V_{Tb}^{\mathrm{thr}}$\ (%
\ref{V_thr}) is overcomed (Fig. 7). This fast increase is caused by the fast
demagnetization of the beam ions, in which case they cannot compensate the
perturbed currents of fully magnetized electrons. In the well over-threshold
regime $\alpha _{b}>$\ $3\alpha _{b}^{\mathrm{thr}}$\ the temperature
dependence weakens because of the nearly saturated demagnetization.

4. From the growth rate $\gamma _{\mathrm{max}}$ (\ref{gm}) it follows that
the instability can be strong, $\gamma _{\mathrm{max}}\gtrsim $\ $0.1\omega
_{Bi}$, even for modest $\alpha _{b}\lesssim $\ $2\alpha _{b}^{\mathrm{thr}}$
not far from the threshold. The most unstable wavenumber $k_{z}\rho
_{Tb}\gtrsim $ $1.54$ near the threshold $\alpha _{b}\gtrsim $ $\alpha _{b}^{%
\mathrm{thr}}$, but increases with $\alpha _{b}$ quickly approaching the
asymptotic scaling $k_{z}\rho _{Tb}\sim $ $\alpha _{b}/2$. In this
asymptotic regime, our growth rate reduces to (\ref{11d}), the same as was
obtained by \cite{Bell2004}.

5. Two particular applications to the terrestrial foreshocks and supernova
remnants show that the reactive CCPI can operate there. Analysis of Section
5.2 suggests that the ions reflected from the shocks around supernova
remnants can drive stronger instability than the cosmic rays. In the
terrestrial foreshock, CCPI can regulate beam parameters generating
electromagnetic fluctuations observed at $k_{z}\rho _{Ti}\approx $\ $0.1$.

Our results complement and extend previous studies on electromagnetic
instabilities and their role in space and astrophysical plasmas. CCPI can
develop around supernova remnants expanding into interstellar medium,
participating in the braking process, heating and redistributing energy in
the supernova shocks. The same concerns the solar-wind regions upstream of
the terrestrial bow shock, as well as other heliospheric shocks, where CCPI
can bound the beam parameters and contribute to the low-frequency
electromagnetic turbulence. Similarly, CCPI can affect other space and
astrophysical environments containing super-Alfv\'{e}nic ion beams and
return currents.

\section*{Acknowledgments}

This research was supported by the Belgian Science Policy Office (through
Prodex/Cluster PEA 90316 and IAP Programme project P7/08 CHARM).

\end{document}